\documentclass[preprint,showpacs,preprintnumbers,superscriptaddress,aps]{revtex4}
\usepackage[colorlinks,linkcolor=blue,anchorcolor=blue,citecolor=blue]{hyperref}
\usepackage{amsmath}
\usepackage{amssymb}
\usepackage{amsfonts}
\usepackage{makecell} 
\usepackage{graphicx}
\usepackage{dcolumn}
\usepackage{bm}

\makeatletter
\def\@pacs@print{} 
\def\pacs#1{\def\@pacs{#1}} 
\makeatother
\begin{document}
\title{A theoretical model for quantifying the imprinting sensitivity of direct-drive inertial confinement fusion implosions} 
\small
\author{Dongxue Liu}
\affiliation{Shanghai Institute of Laser Plasma, Shanghai 201800, China
}
\author{Jiaqin Dong}
\email[e-mail:]{dongjiaqin@hotmail.com}
\affiliation{Shanghai Institute of Laser Plasma, Shanghai 201800, China
}
\author{Yunxing Liu}
\affiliation{Shanghai Institute of Laser Plasma, Shanghai 201800, China
}
\author{Zhiyu He}
\affiliation{Shanghai Institute of Laser Plasma, Shanghai 201800, China
}
\author{Wei Wang}
\affiliation{Shanghai Institute of Laser Plasma, Shanghai 201800, China
}
\author{Yuqiu Gu}
\affiliation{Shanghai Institute of Laser Plasma, Shanghai 201800, China
}
\author{Xiuguang Huang}
\affiliation{Shanghai Institute of Laser Plasma, Shanghai 201800, China
}
\author{Jian Zheng}
\affiliation{Department of Plasma Physics and Fusion Engineering, and CAS Key Laboratory of Frontier Physics in Controlled Nuclear Fusion, University of Science and Technology of China, Hefei 230026,  China
}
\affiliation{Collaborative Innovation Center of IFSA, Shanghai Jiao Tong University, Shanghai, 200240, China}

\date{\today}

\begin{abstract}
To quantify the sensitivity of diverse implosion designs to laser imprinting, we developed an equivalent perturbation model that maps laser imprinting as the initial target surface perturbation. By incorporating imperfections in target fabrication and thermal smoothing in the plasma, the model shows a reduced implosion sensitivity to laser imprinting, extending the analysis beyond geometric irradiation. The imprinting sensitivity threshold is defined as $\frac{\delta h_{\text{proxy}}}{\delta h_{\text{tar}}(0)} = 0.1$, where $\delta h_{\text{proxy}}$ is the imprinting amplitude and $\delta h_{\text{tar}}(0)$ is the initial target perturbation amplitude. Radiation-hydrodynamics simulations confirm that when $\frac{\delta h_{\text{proxy}}}{\delta h_{\text{tar}(0)}} \leq 0.1$, variations in nonlinear onset time and adiabat remain within 12\% of that with $\delta h_{\text{tar}}(0)$ alone. Moreover, the imprinting sensitivity is supported by OMEGA experiments.  Overall, for linear perturbations of medium-to-high modes in direct-drive, the model enhances our physical understanding of how laser and target perturbations evolve and serves as a simplified tool to optimize implosion performance.

\hspace{-10pt}Keywords: implosion sensitivity, laser imprinting, equivalent perturbation model, direct-drive 
\end{abstract}

\maketitle 
\section{Introduction}
The achievement of ignition \cite{abu2022lawson,kritcher2022design,abu2024achievement} at the National Ignition Facility (NIF) marks a historic milestone in inertial confinement fusion (ICF). However, the intermediate conversion of laser energy to X‑rays in these indirect‑drive experiments limits high‑gain ignition. Direct‑drive ICF \cite{RN506} represents a more efficient laser–target coupling path to high gain, but faces significant challenges early in the implosion. The conduction zone between the critical density surface and the ablation front is insufficient to smooth out spatial nonuniformity in the early deposited laser energy \cite{bodner1974rayleigh}. Consequently, spatial variations in laser intensity \cite{smalyuk2005fourier,goncharov2006early,liu2022mitigating} and inherent target imperfections \cite{pak2023overview} can trigger the Richtmyer–Meshkov (RM) \cite{zhou2025instabilities, lin2026non} instability and seed the Rayleigh–Taylor (RT) instability at the ablation front. Once amplified by RT growth, these perturbations lead to an experimentally observable degradation, i.e., shell deformation and shell thickening due to the elevation of adiabat  \cite{michel2017measurement}.

A detailed study of RT is essential to assess implosion performance and develop mitigation strategies. Single-mode RT perturbations grow exponentially \cite{takabe1985self,RN303,betti1996self} before the low-density bubble region expands linearly in the nonlinear regime \cite{goncharov2002analytical}. Multimode RT perturbations transition from exponential to quadratic growth \cite{sadot2005observation,zhang2018self,zhang2020nonlinear}. Consequently, strategies, such as pulse shaping\cite{dittrich2014design,tao2023laser}, doped targets \cite{fujioka2004suppression,zheng2022optimizing}, and foam coated targets \cite{hu2018mitigating}, have been proposed to reduce the linear growth rate to delay the nonlinear transition. However, suppressing RT linear growth rate alone does not guarantee a stable implosion \cite{li2025achieving} due to the seeding of multiple perturbations.

Substantial research has been devoted to understanding and mitigating the seeding of perturbations. Various target imperfections \cite{igumenshchev2016three,miller2022instability,lei2024nonlinear,liu2025three,zhou2025instabilities,pokornik2025investigating}, which are pre‑existing, grow and undergo damped oscillations via the ablative RM. In contrast, laser imprinting continues until $kD_\text{ac}>1$ and may undergo damped oscillations \cite{mikaelian2005richtmyer,goncharov2006early,aglitskiy2010basic}, where $k$ is the perturbed wavenumber and $D_\text{ac}$ represents the width of the conduction zone. To mitigate imprinting, beam-smoothing techniques such as distributed phase plates (DPPs) \cite{kato1982random,kato1984random}, smoothing by spectral dispersion (SSD) \cite{skupsky1983uniformity,skupsky1989improved}, and polarization smoothing (PS) with distributed polarization rotators \cite{boehly1999reduction,tsubakimoto1992suppression} have been developed. In SSD-smoothed, kilojoule-class experiments on OMEGA \cite{patel2023effects}, the performance of high-adiabat implosions saturates
even when the root‑mean‑square (rms) of laser deposition remains above the 1\%, a sensitivity threshold reported by Skupsky et al \cite{skupsky1983uniformity}. This implies that target imperfections and plasma smoothing can reduce the imprinting sensitivity, thus motivating a general model to quantify the imprinting sensitivity in diverse implosion designs.
 
In the paper, we developed an equivalent perturbation model that treats laser-imprinted perturbations ($\delta h_\text{proxy}$) as target surface perturbation ($\delta h_\text{tar}(0))$. The model reveals a transition in the slope of the curve, that plots the shift in nonlinear onset time caused by laser imprinting against $\frac{\delta h_\text{proxy} }{\delta h_\text{tar}(0)}$. This transition defines the imprinting sensitivity threshold as $\frac{\delta h_\text{proxy} }{\delta h_\text{tar}(0)}=0.1$. Radiation-hydrodynamics simulations validate that when $\frac{\delta h_\text{proxy} }{\delta h_\text{tar}(0)}\leq0.1$, the variation in the nonlinear onset time and adiabat at the start of the acceleration phase remains within 12\%, compared to simulations with only $\delta h_\text{tar}(0)$. Moreover, the imprinting sensitivity is also supported by OMEGA experiments. These results indicate that joint control of laser and target perturbations can support stable implosions for high-gain direct-drive fusion.

The paper is organized as follows. In Section II, an equivalent perturbation model is developed to quantify laser imprinting amplitude and imprinting sensitivity in various implosion designs. In Section III, the model is validated using radiation-hydrodynamic simulations. The imprinting sensitivity is supported by OMEGA experiments and suggests joint control of laser and target perturbations. In Section IV, we draw our conclusions.
\section{Equivalent perturbation model}
In direct‑drive ICF, laser nonuniformity is characterized by $\delta I_L/I_L$, and target surface perturbation is quantified as $\delta h_\text{tar}(0)$, where $I_L$ represents the laser power. By incorporating $\delta h_\text{tar}(0)$ and plasma smoothing, we develop an equivalent perturbation model to map laser imprinting as the corresponding target surface perturbation. Thus, the model can quantify the implosion sensitivity to laser imprinting, extending the analysis beyond geometric irradiation.
\subsection{Role of RM and RT instabilities}
$\delta I_L/I_L$ and $\delta h_\text{tar}(0)$ can seed the RT instability through the RM instability. To quantify the effect of $\delta I_L/I_L$, we use the RM process to equate laser‑imprinted perturbations, $\delta h_\text{proxy}$, with $\delta h_\text{tar}(0)$, and then use the RT growth as a measure of sensitivity to laser imprinting.

We define the equivalence at a specific time \(t_{\text{load}}\), when \(k D_{ac} = 1\). 
After \(t_{\text{load}}\), both types of perturbations evolve together under the ablative RM and RT instabilities. Our equivalent condition is therefore expressed as
\begin{equation}
 \delta h_{\text{las}}(t_{\text{load}}) = \delta h_{\text{tar}}(t_{\text{load}}).   
\end{equation} To separate the initial perturbations from their growth, we rewrite this condition as,
\begin{equation}
 \delta h_{\text{proxy}} + \delta_1 = \delta h_{\text{tar}}(0) + \delta_2.   
\end{equation}
Here, \(\delta h_{\text{proxy}}\) is the laser perturbation at \(t_{\text{load}}\) excluding the ablative RM growth \(\delta_1\) when \(k D_{ac} <1\). Similarly, \(\delta_2\) is the growth of \(\delta h_{\text{tar}}(0)\) over the same interval. In practice, we use the simpler proxy,
\begin{equation}
 \delta h_{\text{proxy}} = \delta h_{\text{tar}}(0),   
\end{equation}
to represent the laser imprinting as the target surface perturbation. This approximation introduces an error because \(\delta_1\) and \(\delta_2\) are generally not equal, and their difference, \(\delta_2 - \delta_1\), grows with \(t_{\text{load}}\). Nevertheless, \(\delta h_{\text{proxy}}\) is a convenient scaling parameter for quantifying laser imprinting, since predicting \(\delta h_{\text{las}}(t_{\text{load}})\) or \(\delta h_{\text{tar}}(t_{\text{load}})\) directly is difficult. For typical direct‑drive designs where a prepulse is followed by a main pulse, the model works well for medium‑to‑high modes \cite{hurricane2023physics}. This is because laser imprinting occurs during the short prepulse stage, making \(t_{\text{load}}\) small.
 
 Implosion performance is highly sensitive to RT. For linear $\delta h_{\text{tar}}(0)$, laser imprinting does not alter the linear growth rate; instead, it affects the transition into the nonlinear regime by modifying the initial perturbation amplitude at the onset of the acceleration phase. Therefore, we define the sensitivity of laser imprinting with respect to the nonlinear onset of RT. The perturbation seeds at the start of acceleration are \(\delta h_{\text{las}}(t_\text{on})\) and \(\delta h_{\text{tar}}(t_\text{on})\), which differ from \(\delta h_{\text{proxy}}\) and \(\delta h_{\text{tar}}(0)\). According to Goncharov’s theory of ablative RM instability \cite{goncharov2006early} at a corrugated laser-facing interface, during the shock‑loading phase of a direct‑drive thin‑shell design, the ablation front is still in its first oscillation. This oscillation has a frequency \(\omega = k \sqrt{v_a v_{bl}}\) and an amplitude that decays as \(e^{-k v_a}\), where \(v_a\) and \(v_{bl}\) are the ablative velocity and blow‑off velocity, respectively. Consequently, for perturbations of the same scale, the following ratio holds:
\begin{equation}
 \frac{\delta h_{\text{las}}(t_\text{on})}{\delta h_{\text{tar}}(t_\text{on})} = \frac{\delta h_{\text{proxy}}}{\delta h_{\text{tar}}(0)}.   
\end{equation}
Therefore, we simply use the ratio \(\delta h_{\text{proxy}} / \delta h_{\text{tar}}(0)\) as our measure of imprinting sensitivity.

\subsection{Perturbation seeding process before $t_\text{load}$}
\begin{figure}[h] 
  \begin{minipage}{0.47\textwidth}
(a)\includegraphics[width=0.9\linewidth]{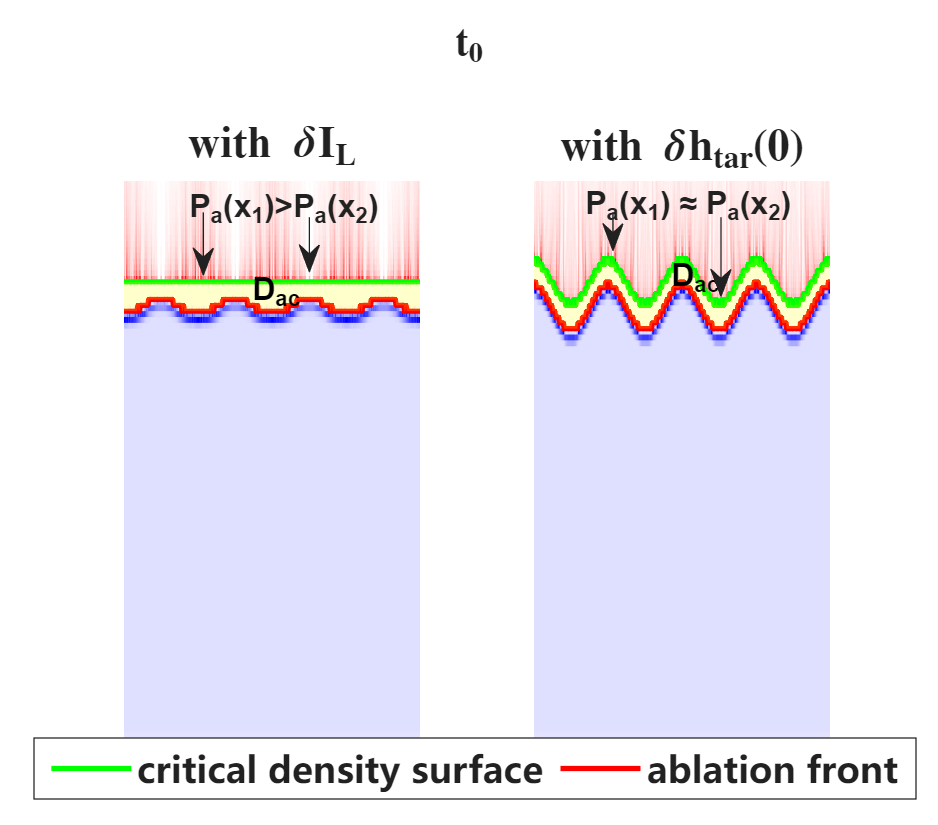}%
\end{minipage}
\begin{minipage}{0.47\textwidth}
(b)\includegraphics[width=0.9\linewidth]{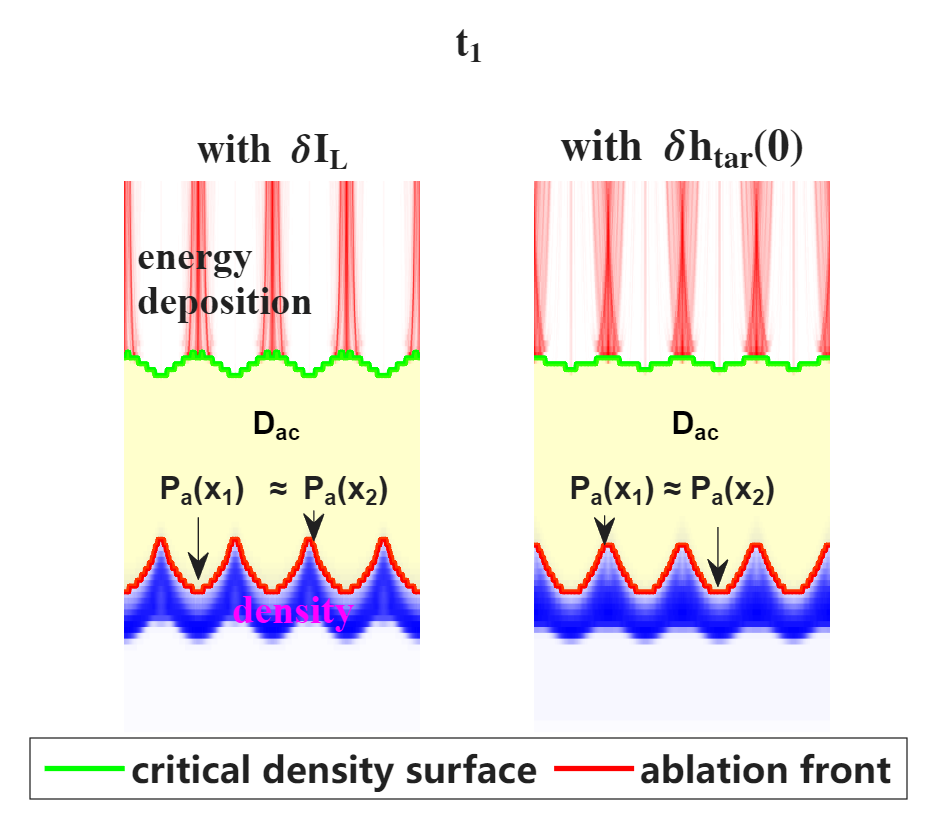}%
\end{minipage}
 \caption{\label{FIG. 1} Illustration of the seeding process for $\delta I_L$ and $\delta h_{\text{tar}}(0)$, both of which are in phase (i.e., a phase difference of 0). (a) Early in irradiation, left: the conduction zone, denoted by the yellow shaded area, is too short to smooth the laser-imprinted pressure perturbation $\delta P_a = P_a(x_1) - P_a(x_2)$. In contrast, right: the ablative pressure $P_a$ is nearly uniform at the ablation front perturbed by $\delta h_{\text{tar}}(0)$, except for slight modulations caused by ray deflection. (b) Later at $t_1$, the extended conduction zone smooths $\delta P_a$ caused by $\delta I_L$ (left) and ray deflection (right) toward high-density regions. In both panels, the blue and red represent the spatial distribution of normalized target density and laser energy deposition, respectively, with lower opacity corresponding to lower values. Detailed simulation setup can be found in the Appendix A.} 
 \end{figure}
To clarify the different seeding processes before $t_\text{load}$,
Figs. \ref{FIG. 1}(a, b) illuminate the seeding of $\delta I_L$ and $\delta h_\text{tar}(0)$, both of which are in phase (i.e., a phase difference of 0). 
For \(\delta I_L\), the short conduction zone \( D_\text{ac}\) cannot smooth the laser deposition variations in the left of Fig. \ref{FIG. 1}(a). This modulates ablative pressure, producing fluctuations in ablative velocity that deform the ablation front in the left of Fig. \ref{FIG. 1}(b). Thus, laser imprinting depends not only on $\delta I_L$, but also on the ablative history. In contrast, for \(\delta h_{\text{tar}}(0)\), the ablative pressure remains nearly uniform on the perturbed ablation front. This is because the energy deposition is initially uniform in the right of Fig. \ref{FIG. 1}(a), and later perturbed energy deposition caused by ray deflection is smoothed by the extended conduction zone in the right of Fig. \ref{FIG. 1}(b). Thus, the deformation of the ablation front can be traced directly to \(\delta h_{\text{tar}}(0)\). Note that in the left of Fig. \ref{FIG. 1}(b), the higher laser energy deposition increases the mass ablation rate, which in turn causes a local depression of the target. Therefore, when $\delta I_L$ is in phase with $\delta h_\text{tar}(0)$, the laser imprinted perturbation (the left of Fig. \ref{FIG. 1}(b)) becomes antiphase (i.e., a phase difference of $\pi$) with $\delta h_\text{tar}(0)$ (the right of Fig. \ref{FIG. 1}(b)).
\subsection{Model derivation }

Building on the seeding process, we developed an equivalent perturbation model for laser perturbations that satisfy \(k D_{ac}(t) \geq 1\) early in irradiation. The model relates the imprinted laser perturbation to the equivalent target surface perturbation. 

Following the cloudy-day model \cite{bodner1974rayleigh}, pressure nonuniformity with wavenumber \(k\) decays exponentially as \(e^{-kD_{ac}(t)}\). Under steady ablation of polyethylene, the ablative pressure and velocity are given by \cite{dahmani1991laser}:
\begin{equation}
P_a = 11.9 \left( I_L / \lambda_L \right)^{2/3} \quad \text{[Mbar]},
\label{eq:pressure}
\end{equation}
\begin{equation}
v_a = 3.08 \times 10^5 \left( I_L / \lambda_L^4 \right)^{1/3} /\rho_a \quad \text{[cm/s]},
\label{eq:velocity}
\end{equation}
where the laser power \(I_L\) is in \(10^{14}\text{W/cm}^2\), the density at the ablation front \(\rho_a\) is in \(\text{g/cm}^3\), and the laser wavelength \(\lambda_L\) is in \(\mu\text{m}\). The corresponding nonuniformity in ablative velocity decays as \(e^{-kD_{ac}(t)/2}\), yielding:
\begin{equation}
\begin{aligned}
   \delta v_{a+}(t) &= v_a \left( \left(1 + \frac{\delta I_L}{I_L}\right)^{1/3} - 1^{1/3} \right)  e^{-k D_{ac}(t)/2}  \quad \text{[cm/s]}, \\
   \delta v_{a-}(t) &= v_a \left(1 ^{1/3} -\left(1 -\frac{\delta I_L}{I_L}\right)^{1/3}  \right)  e^{-k D_{ac}(t)/2}  \quad \text{[cm/s]},
\end{aligned}
\label{eq:velocity_perturbation}
\end{equation}
where \(\delta v_{a+}(t)\) and \(\delta v_{a-}(t)\) are the maximum and minimum perturbations in ablative velocity. The linear deformation of the ablation front is obtained by integrating \(\delta v_a(t)\) when the conduction zone remains insufficient to smooth the laser imprinting, i.e. \(k D_{ac}(t) \leq 1\):
\begin{equation}
\delta h_{\text{proxy}} = \frac{1}{2}\left(\int_{k D_{ac}(t) \leq 1} \delta v_{a+}(t) dt +\int_{k D_{ac}(t) \leq 1} \delta v_{a-}(t) dt\right)\quad \text{[cm]}.
\label{eq:deformation}
\end{equation}
Here, $\frac{\delta h_{\text{proxy}}}{\delta h_{\text{tar}}(0)} = 1$ at $kD_{\text{ac}} = 1$ implies that the laser imprinting amplitude is equivalent to the target surface perturbation amplitude. For typical direct-drive pulse designs, the prepulse leads to a rapid increase in the ablative velocity and a fast growth of the conduction zone. Therefore, the time variation in $kD_{ac} (t)\leq1$ and $kD_{ac} (t)\leq2$ is small. 
The equivalence is not a simple linear function of $\delta I_L$; it depends critically on the laser pulse and the perturbed wavelength through $\delta v_a$. This dependency is detailed in the Table \ref{T1} in the appendix. 
\subsection{Quantifying imprinting sensitivity}
Further analysis on OMEGA experiments \cite{patel2023effects} demonstrate that $\delta h_{\text{tar}}(0)$ and plasma smoothing can reduce the implosion sensitivity to laser imprinting\cite{cao2023impact}. Using the equivalent perturbation model that incorporates the effects of both $\delta h_{\text{tar}}(0)$ and plasma smoothing, we quantify $\frac{\delta h_{\text{proxy}}}{\delta h_{\text{tar}}(0)}$ and derive the resulting shift in nonlinear onset time due to laser imprinting. The dependence of the shift on $\frac{\delta h_{\text{proxy}}}{\delta h_{\text{tar}}(0)}$ determines the imprinting sensitivity. When the implosion is insensitive to laser imprinting, it resides in a target‑dominant regime governed by $\delta h_{\text{tar}}(0)$.

We analyze how the coexistence of $\delta h_{\text{proxy}}$ and $\delta h_{\text{tar}}(0)$ affects the nonlinear onset time, compared to the scenario with only $\delta h_{\text{tar}}(0)$. When only $\delta h_{\text{tar}}(0)$ is present, the nonlinear onset time for a single‑mode perturbation is defined as the time when its amplitude reaches $0.1\lambda$ \cite{haan1989onset}. For a reference perturbation with $\delta h_{\text{ref}}=2\;\mu\text{m}$ and $\lambda_{\text{ref}}=50\;\mu\text{m}$ at the start of acceleration, the reference onset time $t_{\text{ref}}$ is:
 \begin{equation}
  t_{\text{ref}} = \frac{1}{\gamma_{50}} \ln\left( \frac{5}{2 } \right),
 \label{eq:ref_time}
 \end{equation}
 where $\gamma_{50}$ is the linear growth rate of $\lambda_{\text{ref}}$.  When $\delta h_{\text{proxy}}$ and $\delta h_{\text{tar}}(0)$ with the same wavelength coexist, the perturbation amplitude is modified. As the phase difference between $\delta h_{\text{proxy}}$ and $\delta h_{\text{tar}}(0)$ ranges from $0$ (in‑phase) to $\pi$ (antiphase), compared to the case with only $\delta h_{\text{tar}}(0)$, the total amplitude is scaled by a factor between $\frac{|\delta h_{\text{proxy}} - \delta h_{\text{tar}}(0)|}{\delta h_{\text{tar}}(0)}$ and $\frac{\delta h_{\text{proxy}} + \delta h_{\text{tar}}(0)}{\delta h_{\text{tar}}(0)}$. In particular, when $\delta h_{\text{proxy}}$ is antiphase with $\delta h_{\text{tar}}(0)$, the total amplitude is reduced, which delays the nonlinear onset time,

\begin{equation}
  t_{\text{lt}} -  t_{\text{t}} =
\frac{-\ln\!\left(1-\frac{\delta h_{\text{proxy}}}{\delta h_{\text{tar}}(0)}\right)}
{\ln\!\bigl(5/2\bigr)} \frac{\gamma_{50}}{\gamma_{\lambda}}\; t_{\text{ref}}.
\label{eq:coupling_shift1}
\end{equation}
Here, \( t_{\text{lt}}\) and \( t_{\text{t}}\) denote the nonlinear onset time with both perturbations and with only $\delta h_{\text{tar}}(0)$, respectively, $\gamma_{\lambda}$ represents the linear growth rate of the single-mode perturbation, and the range of $\frac{\delta h_{\text{proxy}}}{\delta h_{\text{tar}}(0)}$ is zero to one. Conversely, when $\delta h_{\text{proxy}}$ and $\delta h_\text{tar}(0)$ are in-phase, the nonlinear onset time advances:
\begin{equation}
  t_{\text{lt}} -  t_{\text{t}} =
\frac{-\ln\!\left(1+\frac{\delta h_{\text{proxy}}}{\delta h_{\text{tar}}(0)}\right)}
{\ln\!\bigl(5/2\bigr)} \frac{\gamma_{50}}{\gamma_{\lambda}}\; t_{\text{ref}}.
\label{eq:coupling_shift2}
\end{equation}  
Generally, when the phase difference between $\delta h_{\text{proxy}}$ and $\delta h_\text{tar}(0)$ lies in the range $(0,\pi)$, the perturbation amplitude is scaled between $\frac{|\delta h_{\text{proxy}} - \delta h_{\text{tar}}(0)|}{\delta h_{\text{tar}}(0)}$ and $\frac{\delta h_{\text{proxy}} + \delta h_{\text{tar}}(0)}{\delta h_{\text{tar}}(0)}$, and the shift in the onset time of nonlinearity falls between Eq. \ref{eq:coupling_shift2} and Eq. \ref{eq:coupling_shift1}.

\begin{figure}[h]
\begin{minipage}{0.47\textwidth}
(a)\includegraphics[width=0.9\linewidth]{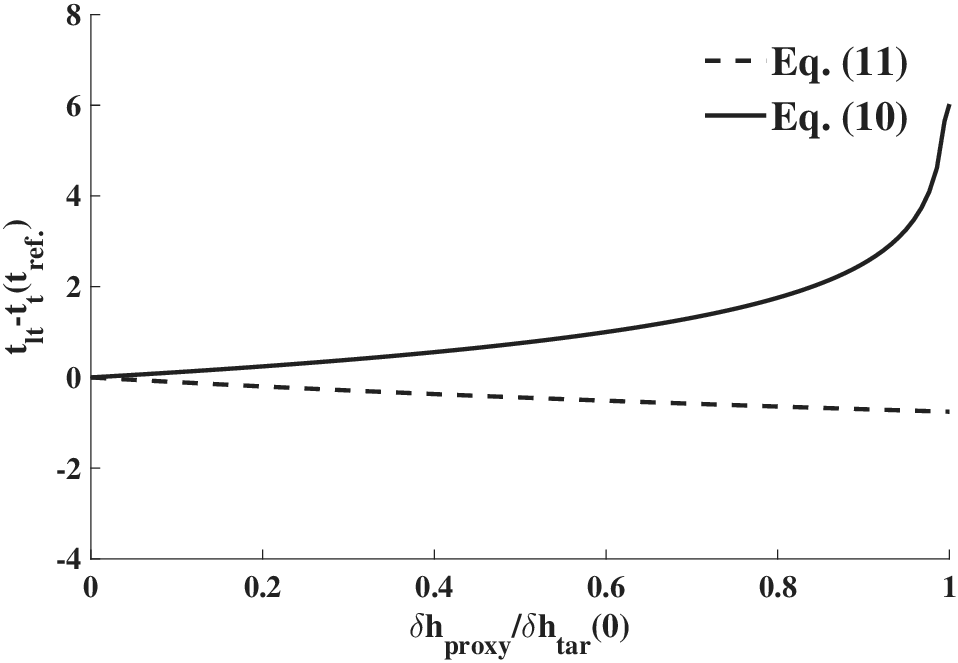}%
 \end{minipage}
\begin{minipage}{0.47\textwidth}
(b)\includegraphics[width=0.9\linewidth]{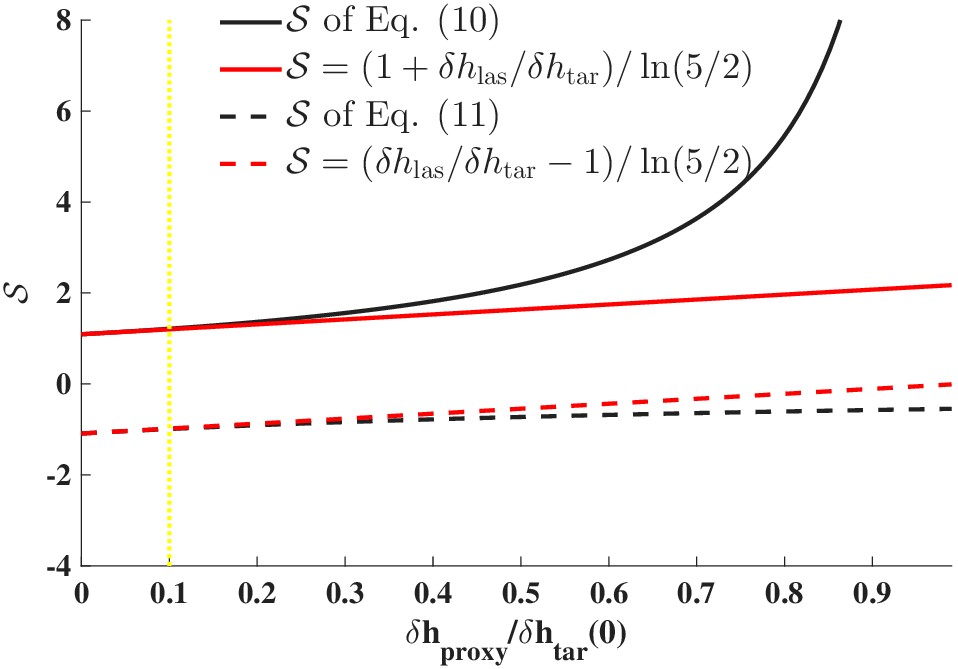}%
 \end{minipage}

  \caption{\label{FIG. 2} Evolution of (a) \(t_{\text{lt}}-t_{\text{t}}\) and (b) $\mathcal{S}=\frac{d(t_\text{lt}-t_\text{t})}{d(\delta h_\text{proxy}/\delta h_\text{tar}(0))}$ as a function of  $\frac{\delta h_\text{proxy}}{\delta h_\text{tar}(0)}$. In (a), the solid line shows the delay when $\delta h_\text{proxy}$ mitigates $\delta h_\text{tar}(0)$; the dashed line shows the advancement caused by $\delta h_\text{proxy}$.
 In (b),  $\mathcal{S}$ shifts from linear to nonlinear growth as increasing $\frac{\delta h_\text{proxy}}{\delta h_\text{tar}(0)}$. The onset of this transition, marked by the yellow dotted line, occurs at $\frac{\delta h_\text{proxy}}{\delta h_\text{tar}(0)}=0.1$. The red lines fit the linear growth of $\mathcal{S}$. } 
 \end{figure}

Fig. \ref{FIG. 2}(a) plots Eq. \ref{eq:coupling_shift1} and Eq. \ref{eq:coupling_shift2} for $\gamma_{\lambda}=\gamma_{50}$.  $t_{\text{lt}} -  t_{\text{t}}$ approaches zero for the weak laser imprinting amplitude. The slope $\mathcal{S}=\frac{d(t_\text{lt}-t_\text{t})}{d(\delta h_\text{proxy}/\delta h_\text{tar}(0))}$ measures how rapidly the nonlinear onset time changes with imprinting. Fig. \ref{FIG. 2}(b) shows $\mathcal{S}$ as a function of \(\frac{\delta h_{\text{proxy}}}{\delta h_{\text{tar}}(0)}\). Its behavior reveals a clear transition: $\mathcal{S}$ increases linearly for \(\frac{\delta h_{\text{proxy}}}{\delta h_{\text{tar}}(0)}\leq 0.1\) but grows nonlinearly thereafter. This transition defines the imprinting sensitivity threshold:
\begin{equation}
\frac{\delta h_{\text{proxy}}}{\delta h_{\text{tar}}(0)}= 0.1 .
\label{condition}
\end{equation}
A target-dominant regime is then characterized by $\frac{\delta h_{\text{proxy}}}{\delta h_{\text{tar}}(0)} \leq 0.1 $. Within this regime, the relative shift in the nonlinear onset time, \((t_{\text{lt}}-t_{\text{t}})/t_{\text{ref}}\), remains below 12\%. Importantly, Eq. \ref{condition} should be understood as an empirical, design-oriented metric, whose specific value is tied to the linear or weakly nonlinear regime below the ignition cliff. In practice, the tolerance should be chosen according to the design's margin relative to the ignition cliff. For designs closer to the cliff, a tighter tolerance (e.g., 5\%) is advisable, which would lower the corresponding threshold under the influence of strong nonlinearity. Conversely, for designs with greater margin, a looser tolerance (e.g., 15\%) may be acceptable, leading to a higher threshold.
 \subsection{Real application and limitation in multimode}
 The above analysis of equivalence and imprinting sensitivity applies to linear perturbations of single medium‑to‑high modes in direct drive.  
However, realistic implosions involve multimode perturbations, the phases of which evolve over time and vary from mode to mode. This imposes additional limitations on the applicability of the equivalence and imprint sensitivity.

The experimentally measurable root‑mean‑square amplitudes are $\sigma_\text{tar}(0)$ for the target surface perturbation and $\sigma_\text{las}$ for the laser perturbation. For $\sigma_\text{las}$ with a perturbed spectrum over the wavenumber range $[k_1, k_2]$, where smaller wavenumbers produce stronger imprinting, the resulting imprinting amplitude lies between $\delta h_{\text{proxy2}}$ (corresponding to $k_2$) and $\delta h_{\text{proxy1}}$ (corresponding to $k_1$). To ensure that the range $\delta h_{\text{proxy1}} - \delta h_{\text{proxy2}}$ is smaller than the measurement uncertainty, the spectral bandwidth must be sufficiently narrow. This approach avoids the need to identify the dominant perturbed mode.

The phase difference between the linear $\delta h_{\text{proxy1}}$ and $\sigma_{\text{tar}}(0)$ can range from $0$ to $\pi$. Therefore, regardless of how the phase varies in time and space, the total amplitude, when both $\delta h_{\text{proxy1}}$ and $\sigma_{\text{tar}}(0)$ are present, is scaled by a factor between $\frac{|\delta h_{\text{proxy1}} - \sigma_{\text{tar}}(0)|}{\sigma_{\text{tar}}(0)}$ and $\frac{\delta h_{\text{proxy1}} + \sigma{\text{tar}}(0)}{\sigma_{\text{tar}}(0)}$, relative to the case with only $\sigma_{\text{tar}}(0)$. Moreover, to maintain the ratio $\delta h_{\text{las}}(t_\text{on}) / \delta h_{\text{tar}}(t_\text{on}) = \delta h_{\text{proxy1}} / \sigma_{\text{tar}}(0)$, the imprinted laser perturbation must not alter the fastest‑growing mode. Beyond the linear regime, the imprinting sensitivity model fails due to the absence of mode coupling or competition \cite{zhou2025instabilities}, and the nonlinear onset time shift lies outside the region bounded by Eq. \ref{eq:coupling_shift1} and Eq. \ref{eq:coupling_shift2}. 

For multimode, the nonlinear stage begins when the maximum deformation reaches
\( 0.1\lambda_{\text{ran}}=[\lambda_2,\lambda_1]\), where $\lambda_{\text{ran}}$ is the perturbed wavelength range. The determination of the nonlinear onset time is closely related with the  dominated mode. When $\delta h_{\text{proxy}}$ and $\delta h_{\text{tar}}$ are in-phase and the dominant mode is $k1$, the shift in the nonlinear onset time becomes \begin{equation}
  t_{\text{lt}} - t_{\text{t}} =
\frac{-\ln\!\left(1+\dfrac{\delta h_{\text{proxy1}}}{\sigma_{\text{tar}}(0)}\right)}
{\ln\!\bigl(5/2\bigr)} \frac{\gamma_{50}}{\gamma_{k_{\text{1}}}}\;  t_{\text{ref}}.
\label{eq:multimode_shift}
\end{equation}
Here, $\gamma_{k_{\text{1}}}$ represents the linear growth rate of the perturbed mode $k1$. The imprinting sensitivity threshold remains similar. For $\frac{\delta h_{\text{proxy}}}{\sigma_{\text{tar}}(0)}\leq 0.1$ and $\gamma_{\lambda}=\gamma_{50}$, $\frac{t_{\text{lt}}-t_\text{t}}{t_{\text{ref}}}$ remains below 12\%.

\section{Multidimensional validation of the model}
\subsection{Simulation validation of the equivalent perturbation model}

We validate the equivalent perturbation model (Eq. \ref{eq:deformation}) by comparing two distinct hydrodynamic scenarios: one involves $\delta I_L$ and the other involves $\delta h_{\text{tar}}(0)$. Validation cases 1-16 (see Appendix Table \ref{T1})) were performed using the Eulerian radiation-hydrodynamics code FLASH \cite{fryxell2000flash}. Under two different laser pulses, all cases satisfy $\frac{\delta h_{\text{proxy}}}{\delta h_{\text{tar}}(0)} \approx 1$ at $kD_{\text{ac}} \approx 1$, indicating the model's potential applicability in various implosion designs.
\begin{figure}[h]
\begin{minipage}{0.44\textwidth}
(a)\includegraphics[width=0.93\linewidth]{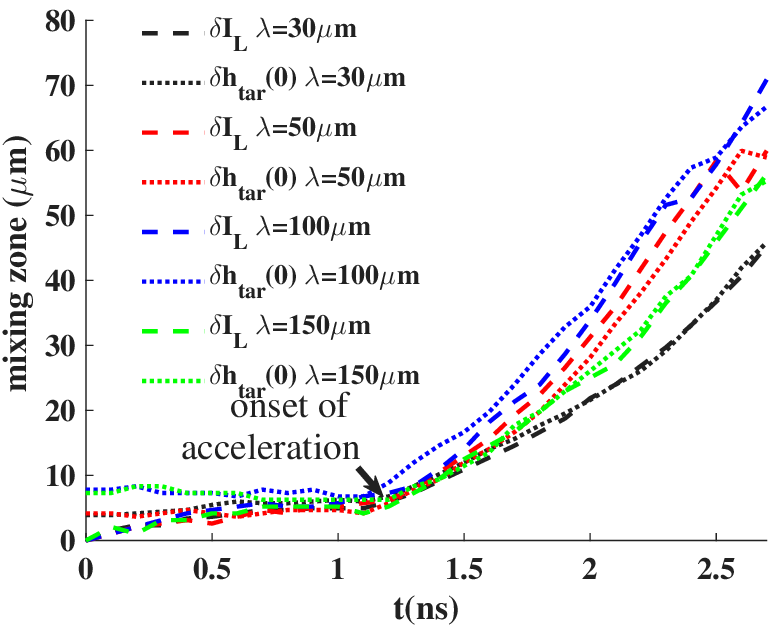}%

(b)\includegraphics[width=0.93\linewidth]{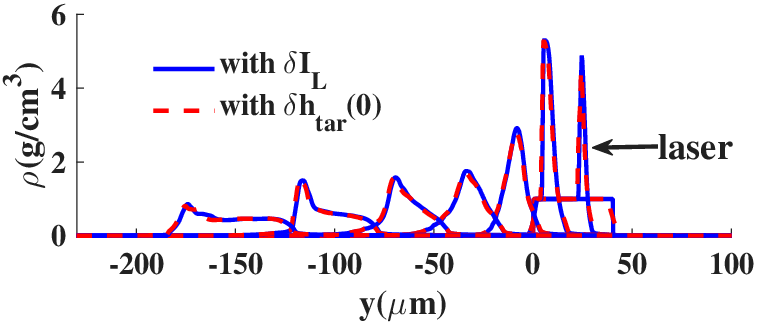}%
 \end{minipage}
\begin{minipage}{0.55\textwidth}
(c)
\includegraphics[width=1.0\linewidth]{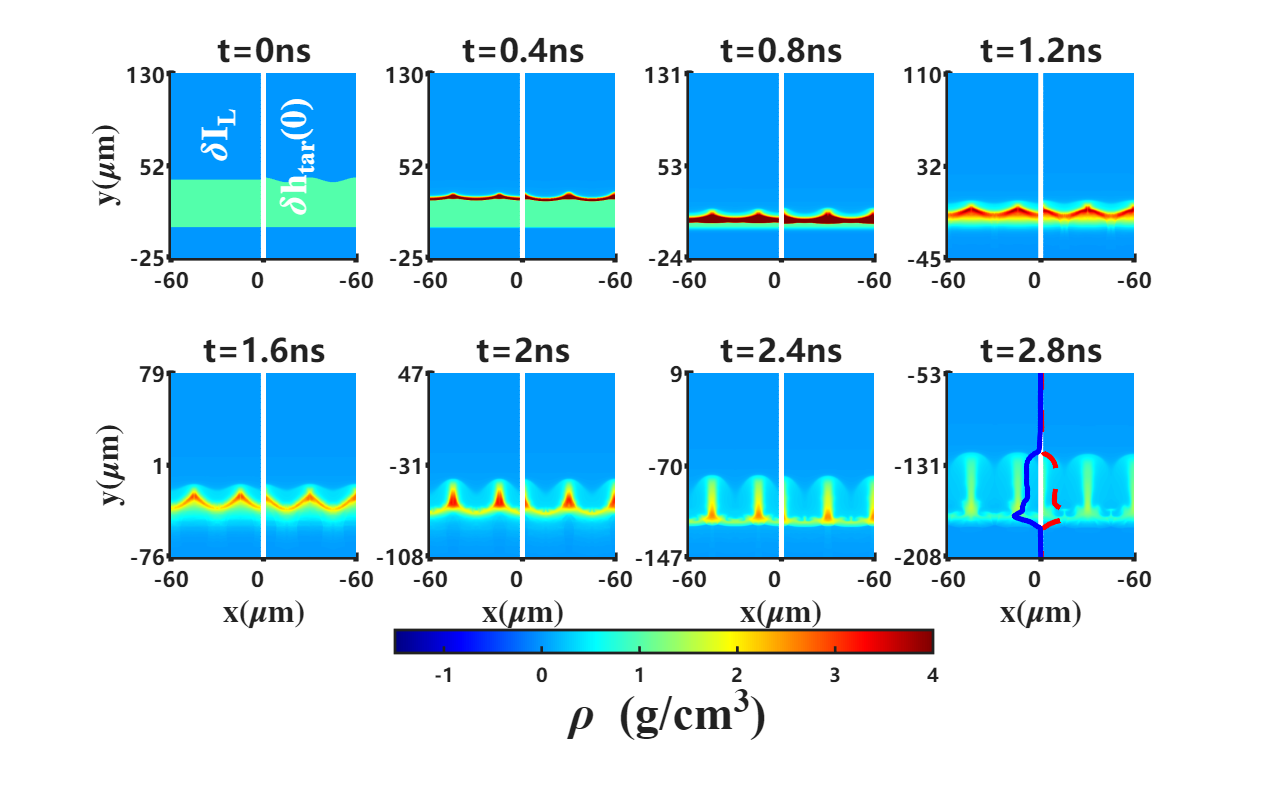}%
\end{minipage}
 \caption{\label{FIG. 3} Temporal evolution of (a) the mixing zone width \(h_\text{mix}\), and (b) the vertically averaged one‑dimensional (1D) and two‑dimensional (2D) density distributions. In (a), the growth of \(h_\text{mix}\) due to $\delta I_L$ (dashed) and $\delta h_{\text{tar}}(0)$ (dotted) at different scales is consistent. \(h_\text{mix}\) is defined as the distance between the maximum and minimum position, where the density at $x=\text{const}$ equals 
$1/e$ of its maximum value near the corona region. In (b), the 1D density profiles for \(\lambda = 30\ \mu\text{m}\) correspond to the red and blue lines at \(t = 2.8\ \text{ns}\) in (c). The left region of the white line represents simulations with \(\delta I_L\); the right region with \(\delta h_{\text{tar}}(0)\). The similarity in \(h_\text{mix}\), 2D, and 1D distributions indicates that the laser‑imprinting amplitude \(\delta h_{\text{proxy}}\) equals the target surface perturbation \(\delta h_{\text{tar}}(0)\). Corresponding simulation cases are 1–8 in the Appendix.} 
  
 \end{figure}

The typical simulation results are shown in Fig. \ref{FIG. 3}. In Fig. \ref{FIG. 3}(a), the mixing zone, $h_\text{mix}$ is induced by $\delta I_L$ (dashed lines) and $\delta h_{\text{tar}}(0)$ (dotted lines) under the same laser pulse. \(h_\text{mix}\) is defined as the distance between the maximum and minimum position, where the density at $x=\text{const}$ equals 
$1/e$ of its maximum value  near the corona region. The evolution of $h_\text{mix}$ is consistent across different scales. Moreover, the one‑dimensional (1D) averaged density profiles for $\lambda = 30\ \mu\text{m}$ in Fig. \ref{FIG. 3}(b) are nearly identical, indicating similarity in the flow field distribution within the mixing zone. The averaging operation is illustrated by the red and blue lines in the two‑dimensional (2D) density distribution shown in Fig. \ref{FIG. 3}(c). Among cases in Fig. \ref{FIG. 3}(a), the mixing zone for $\lambda = 100\ \mu\text{m}$ grows fastest, resulting from the competition between ablative stabilization and the classical RT linear growth rate.  

In Fig. \ref{FIG. 3}(c), the left region of the white line corresponds to simulations with $\delta I_L$, while the right region denotes simulations with $\delta h_{\text{tar}}(0)$. Initially ($t < 0.8\ \text{ns}$), the perturbation amplitude on the right is larger than that on the left. This difference decreases over time and eventually vanishes, consistent with the trends in Fig. \ref{FIG. 3}(a) and Fig. \ref{FIG. 3}(b). These consistencies confirm that the equivalent perturbation model successfully equates laser imprinting to $\delta h_{\text{tar}}(0)$. Nevertheless, detailed differences in density distributions persist, which can be attributed to differences in the seeding process before $t_\text{load}$.
\subsection{Simulation validation for imprinting sensitivity}
When the imprinting amplitude is lower than the implosion sensitivity threshold, expressed as Eq. \ref{condition}, the implosion resides in a target-dominant regime. Radiation‑hydrodynamic simulations confirm that within the target-dominant regime, laser imprinting slightly alters the nonlinear onset time and the adiabat at the start of the acceleration phase. This change is less than 12\% compared to simulations containing only $\delta h_{\text{tar}}(0)$.
 \begin{figure}[h]
 \begin{minipage}{0.4\textwidth}
(a)\includegraphics[width=0.95\linewidth]{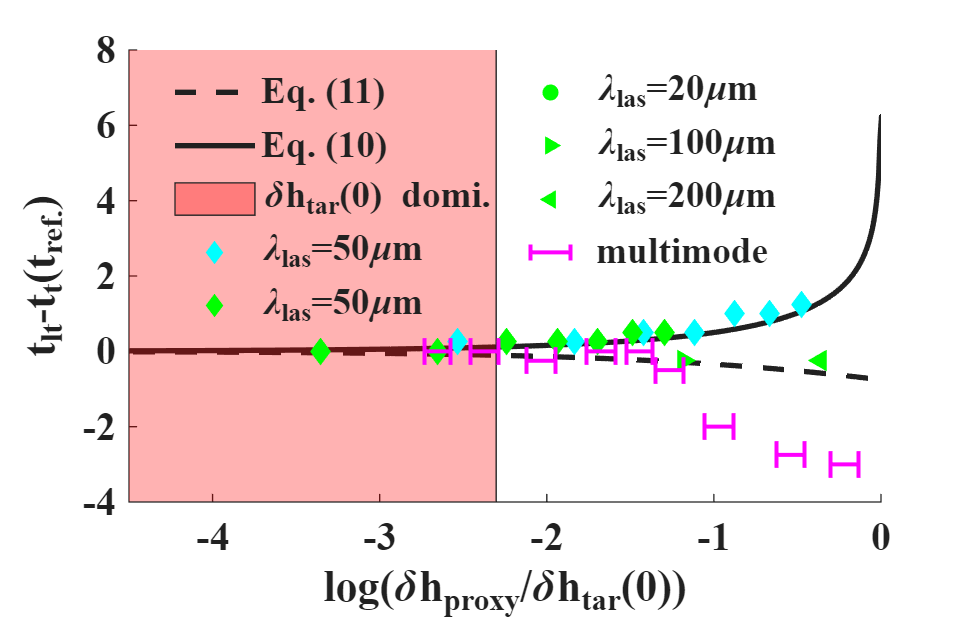}%
 \end{minipage}
 \begin{minipage}{0.58\textwidth}
(b)\includegraphics[width=0.97\linewidth]{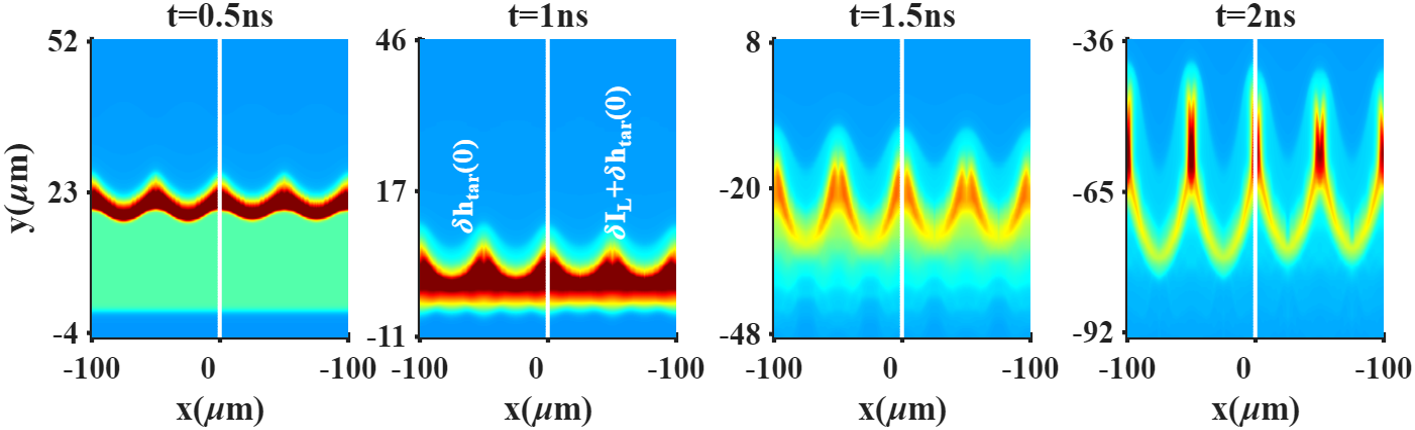}%

(c)\includegraphics[width=0.97\linewidth]{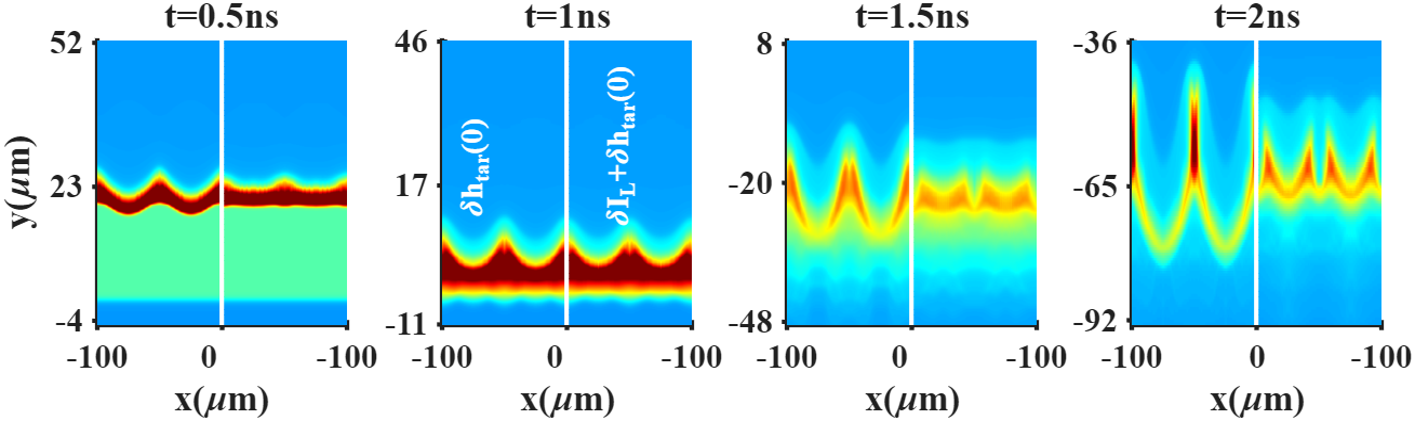}%
 \end{minipage}
 
 \caption{\label{FIG. 4} (a) Evolution of \(t_{\text{lt}}-t_{\text{t}}\) versus $\frac{\delta h_\text{proxy}}{\delta h_\text{tar}(0)}$, plotted on a logarithmic scale and showing same lines in Fig. \ref{FIG. 2}(a). (b) present simulated density profiles at $\frac{\delta h_\text{proxy}}{\delta h_\text{tar}(0)}=0$ (left) and $\frac{\delta h_\text{proxy}}{\delta h_\text{tar}(0)}=0.08$ (right). (c) present simulated density profiles at $\frac{\delta h_\text{proxy}}{\delta h_\text{tar}(0)}=0$ (left) and $\frac{\delta h_\text{proxy}}{\delta h_\text{tar}(0)}=0.62$ (right). Inside the target-dominant regime, dominated  by the red shaded region in (a), laser imprinting has little influence compared with that of $\frac{\delta h_\text{proxy}}{\delta h_\text{tar}(0)}=0$, as evidenced by the similar density distribution in (b). Outside this region, however, laser imprinting strongly affects the perturbation evolution. $\delta h_\text{proxy}$ and $\delta h_\text{tar}(0)$ are set antiphase to show clear differences in (c). In (a), blue and green colors represent $\delta h_\text{tar}(0)=2\mu \text{m}$ and $\delta h_\text{tar}(0)=4\mu \text{m}$, respectively.  $\delta h_\text{tar}(0)$ is replaced by $\sigma_\text{tar}(0)$ for multimode. Each subplot in (b) and (c) has an equal size and the color bar follows that of Fig. \ref{FIG. 3}(a). Corresponding simulation cases are 17–42 in the Appendix.}
 \end{figure}

 Fig \ref{FIG. 4} validates Eq. (\ref{eq:coupling_shift1}) and Eq. (\ref{eq:coupling_shift2}) using cases 17–42, with simulation parameters listed in Table \ref{T2} in the Appendix. Diamonds represent single‑mode perturbations in both \(\delta I_L\) and \(\delta h_{\text{tar}}(0)\) with identical wavelength and phase. Compared with cases having only \(\delta h_{\text{tar}}(0)\), the nonlinear onset time is delayed, consistent with Eq. (\ref{eq:coupling_shift1}). This suggests that the phase difference between \(\delta h_{\mathrm{proxy}}\) and  \(\delta h_{\mathrm{tar}}(0)\) is $\pi$, which can be explained by the phase of $\delta I_L$. Inside the target-dominant regime, dominated  by the red shaded region in Fig. \ref{FIG. 4}(a), laser imprinting has little influence compared with that of $\frac{\delta h_\text{proxy}}{\delta h_\text{tar}(0)}=0$, as evidenced by the similar density distribution in Fig. \ref{FIG. 4}(b). Outside this region, however, laser imprinting strongly affects the perturbation evolution in Fig. \ref{FIG. 4} (c).
 
 Other simulations except diamonds  are considered multi‑mode, including those where the wavelength of \(\delta I_L\) is varied while \(\delta h_{\text{tar}}(0)\) is fixed at 50 \(\mu \text{m}\) (triangles) and those with $\sigma_\text{las}$ and $\sigma_\text{tar}(0)$ (magenta symbols). We define the simulated nonlinear onset time as the moment when the maximum deformation reaches $0.1\lambda$, where 
$\lambda$ is the maximum wavelength of target perturbations. These multi‑mode data agree with Eq. (\ref{eq:coupling_shift2}), which coincides with Eq. (\ref{eq:multimode_shift}). Here, the amplitude of \(\delta h_\text{{proxy}}\) is closely related to the range of perturbed wavelengths; broader spectral widths produce larger horizontal error bars. The multi‑mode results imply that the target perturbation modulates the phase of laser energy deposition, leading to a nearly zero phase difference between \(\delta h_{{proxy}}\) and \(\sigma_{\text{tar}}(0)\).
The multimode tend toward the dashed curve, while the single mode data tend toward the solid curve. The reason for this difference may be multi‑mode effects, but that is beyond the scope of this paper. Nevertheless, the advancement of the nonlinear onset time caused by multi‑mode perturbations remains bounded by Eq. (\ref{eq:coupling_shift1}) and Eq. (\ref{eq:coupling_shift2}). The model still holds for multi‑mode scenarios except three points representing nonlinear perturbations.

\begin{figure}[h]
\includegraphics[width=0.6\linewidth]{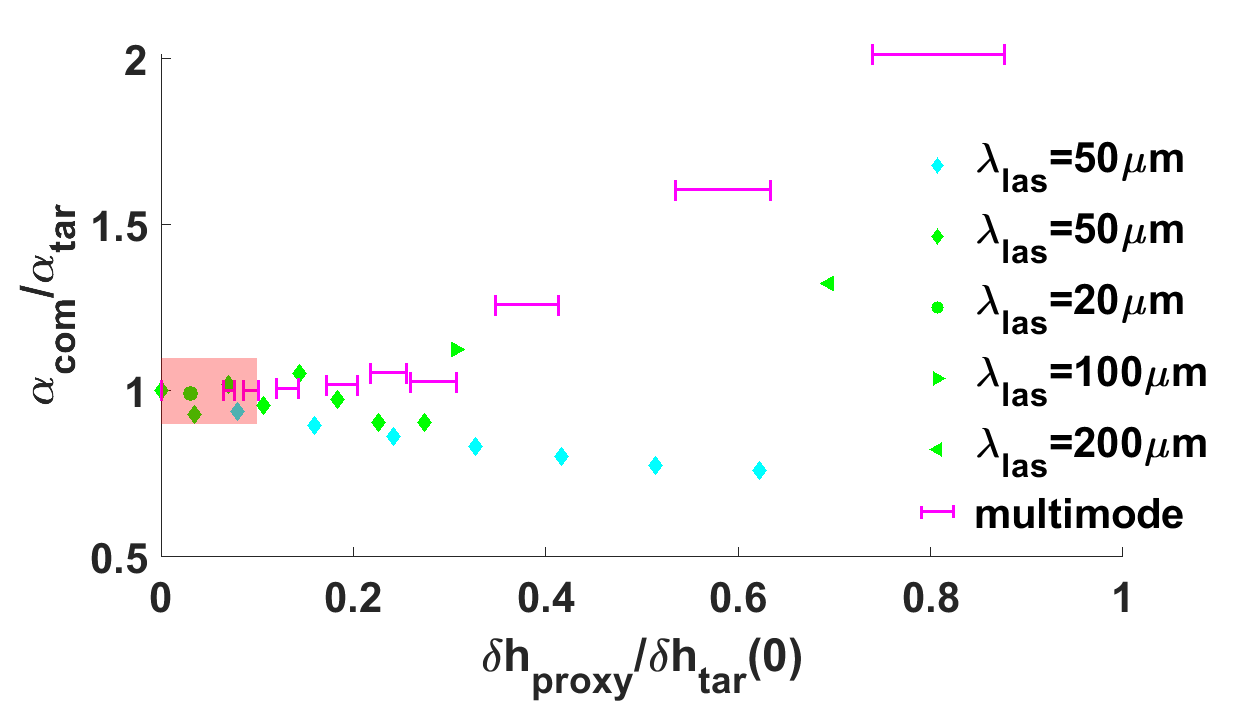}%
 \caption{\label{FIG. 5} Mass weighted adiabat $\frac{\alpha_\text{com}}{\alpha_\text{tar}}\sim \left(\frac{P_\text{com}\rho^3_\text{tar}}{P_\text{tar}\rho^3_\text{com}}\right)_\text{mass}$ versus $\frac{\delta h_\text{proxy}}{\delta h_\text{tar}(0)}$. Symbols follow the same convention as Fig. \ref{FIG. 4}. Within the target‑dominant region (red shaded), laser imprinting alters the adiabat by less than 12\% relative to the case with only $\delta h_{\text{tar}}(0)$.
 } 
 \end{figure}
 
We also evaluate the effect of laser imprinting on the target's mass‑weighted adiabat, expressed as
\begin{equation*}
  \frac{\alpha_\text{com}}{\alpha_\text{tar}}\sim \left(\frac{P_\text{com}\rho^3_\text{tar}}{P_\text{tar}\rho^3_\text{com}}\right)_\text{mass} , 
\end{equation*}
where $\alpha_\text{com}$, $P_\text{com}$ and $\rho_\text{com}$ denote the adiabat, pressure, and density for simulations with both perturbations, and $\alpha_\text{tar}$, $P_\text{tar}$ and $\rho_\text{tar}$ refer to simulations with only  $\delta h_{\text{tar}}(0)$ or $\sigma_{\text{tar}}(0)$. As shown in Fig. \ref{FIG. 5}, within the target‑dominant region, the deviation in adiabat remains below 12\% relative to simulations with only $\delta h_{\text{tar}}(0)$ or ($\sigma_{\text{tar}}(0)$). 

\subsection{Experimental validation for imprinting sensitivity}
The imprinting sensitivity is supported by OMEGA implosion experiments \cite{patel2023effects}. These experiments were designed to systematically probe the dependence of implosion performance on the level of laser imprinting.

On kilojoule-scale cryogenic DT targets, the imprinting level was modulated by varying the SSD bandwidth. The resulting on-target laser nonuniformity $\sigma_{\text{las}}$ is related to the SSD fraction by
\begin{equation}
\sigma_{\text{las}} \approx \frac{20\%}{\sqrt{47.5 \times {SSD_{fraction}} + 1}},
\end{equation}
where ${SSD_{fraction}}$ ranges from 15\% to 100\%. Laser imprinting modes with $l > 20$ dominate the total nonuniformity. The experimental results show that for the high-adiabat design ($\alpha \approx 5$), the neutron yield becomes insensitive to imprinting once $\sigma_{\text{las}}$ falls below approximately 4\%. Here, $\sigma_\text{las}$ represents the total on-target illumination nonuniformity calculated using hard sphere (no plasma) superposition of 60 beam profiles. This threshold is notably lower than the previously reported 1\% \cite{skupsky1983uniformity}. High-dimensional simulations attribute this difference to the combined effects of target perturbations and plasma smoothing.

\begin{table}[h]
    \centering
    \begin{tabular}{c  c  c  c  c  c  c}
    \hline
$\sigma_\text{las}$& 3\% & 4\%
&5\%&6\%&7\%&8\%\\
        \cline{1-7}
         $\sigma_\text{tar}(\mu m)$& 0.4& 0.4&0.4 &0.4&0.4& 0.4\\  
         \cline{1-7}
         $S_{omega}$&N& N & Y&Y&Y&Y\\
         \cline{1-7}
         $\frac{\delta h_\text{proxy}(l=20)}{\sigma_\text{tar}(0)}$&  0.08& 0.1 & 0.13&0.15&0.18&0.20\\
         \hline
    \end{tabular}            
    \caption{Sensitivity of OMEGA implosion experiments to laser imprinting $S_{omega}$, where "N" denotes insensitive and "Y" denotes sensitive. The total on-target illumination nonuniformity $\sigma_\text{las}$ was
calculated using hard sphere (no plasma) superposition of 60 beam profiles, yielding $\sigma_\text{las}\approx20.3(\%)/\sqrt{47.5 \times SSD_{fraction}+1 }$. The amplitude of target perturbations $\sigma_\text{tar}$ corresponds to a $\pm 0.4~\mu$m variation in the CD ablator thickness.} 
    \label{T11}
\end{table}
 
 Our quantitative laser imprinting sensitivity incorporates both $\sigma_\text{tar}(0)$ and plasma smoothing. Stronger plasma smoothing reduces imprinting efficiency, and this reduction, when combined with larger $\sigma_\text{tar}(0)$, decreases the overall implosion sensitivity to $\sigma_\text{las}$. To apply this sensitivity, we performed 1D MULTI simulations \cite{ramis1988multi} using the experimental pulse shape and target design to calculate the time-dependent ablative velocity and conduction zone under non-steady conditions. For the smallest dominant imprinting mode ($l=20$), the perturbation in ablative velocity was derived from Eq. \ref{eq:velocity_perturbation} with $\frac{\delta I_L}{I_L} = \sigma_{\text{las}}$, and the corresponding imprinting amplitude $\delta h_{\text{proxy}}$ was obtained via Eq. \ref{eq:deformation}. The results, summarized in Table \ref{T11}, show that OMEGA implosion performance remained insensitive to laser imprinting when $\delta h_{\text{proxy}} / \delta h_{\text{tar}}(0) \leq 0.10$. Furthermore, for low-entropy pulse designs, $\delta h_{\text{proxy}} / \delta h_{\text{tar}}(0) > 0.10$, the implosion performance increases with the decreasing imprinting level. Those experiments did not reach a burning‑plasma state or approach the ignition cliff; therefore, the observed value is consistent with the threshold based on linear theory. However, once a burning‑plasma state is achieved, the threshold should be modified to match  experimental results. Hence, $\delta h_{\text{proxy}} / \delta h_{\text{tar}}(0)$ preliminarily provides a practical design-oriented metric to assess imprinting sensitivity, which still needs more experimental validation.

 The sensitivity threshold of implosion performance to $\sigma_{\text{tar}}(0)$ can also be derived from the model derivation section, yielding \( \sigma_{\text{tar}}(0)/\delta h_{\text{proxy}}  = 0.10\). This result indicates that for OMEGA implosions using 100\%-bandwidth SSD, $\sigma_{\text{tar}}(0)$ needs to be reduced to 3 $\pm \text{nm}$ to make the implosion performance dominated by $\delta h_{\text{proxy}}$. This requirement exceeds the current capability of target fabrication, making target quality a critical determinant of current implosion performance.
 
\subsection{Implication for joint control of laser and target perturbations}
The quantified sensitivity to laser imprinting, formulated as Eq. \ref{condition}, suggests a practical strategy to jointly control $\delta h_{\text{proxy}}$ and  $\sigma_{\text{tar}}(0)$. When $\delta h_{\text{proxy}} / \sigma_{\text{tar}}(0) \geq 0.1$, implosion performance can be enhanced more effectively by improving laser uniformity. Conversely, when $\delta h_{\text{proxy}} / \sigma_{\text{tar}}(0) <0.1$, greater performance can be achieved by improving target fabrication.

 \begin{figure}[h]
  (a)\includegraphics[width=0.7\linewidth]{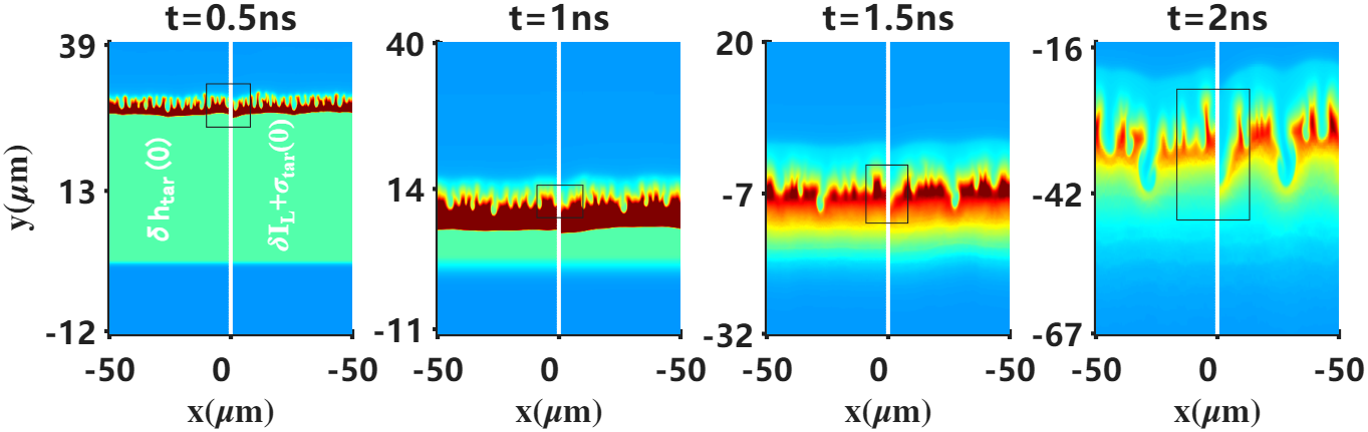}%
  
(b)\includegraphics[width=0.7\linewidth]{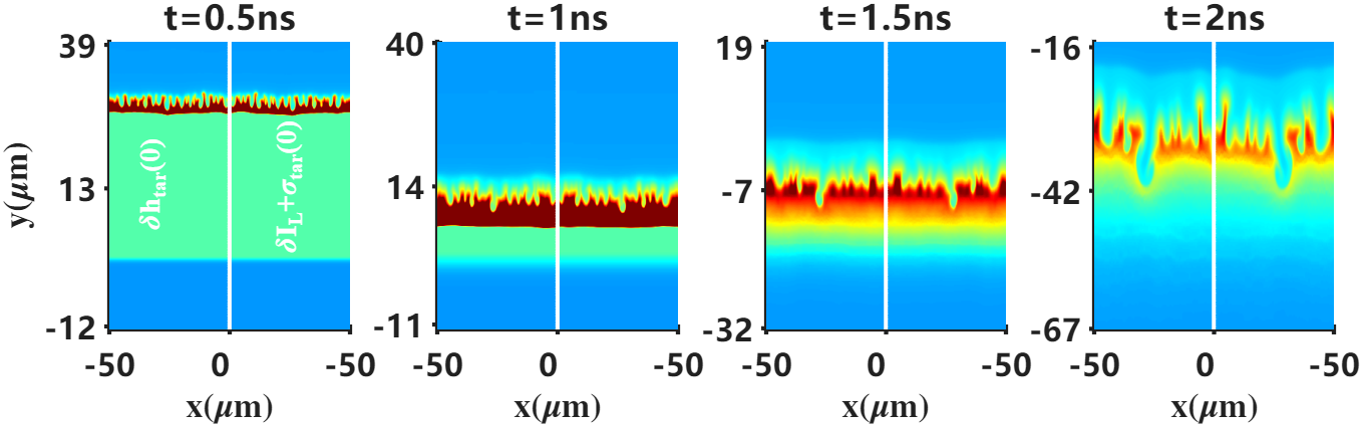}   
 \caption{\label{FIG. 6} Temporal evolution of density under different levels of laser imprinting for multimode perturbations. In (a), the right panel in each subplot has a higher $\frac{\delta h_\text{proxy}}{\sigma_\text{tar}(0)}=0.22\sim 0.26$ and shows clear degradation in shell integrity. To clearly show the difference, the boxed area is placed in the middle. In (b), the right panel in each subplot corresponds to a low imprinting ratio of $\frac{\delta h_\text{proxy}}{\sigma_\text{tar}(0)}=0.07\sim 0.08$, where the shell integrity is similar to the left, $\frac{\delta h_\text{proxy}}{\sigma\text{tar}(0)}=0$. The color bar follows that of Fig. \ref{FIG. 3}(c). The profiles of multimode perturbation are shown in Fig. \ref{FIG. 7}.} 
 \end{figure}
 To demonstrate this strategy, simulations with multimode perturbation were performed and the profiles of multimode perturbation are shown in Fig. \ref{FIG. 7}. The temporal evolution of the shell density under different imprinting levels is shown in Fig. \ref{FIG. 6}. Fig.\ref{FIG. 6}(b) shows that a higher imprinting level (\(\delta h_{\text{proxy}} / \sigma_{\text{tar}}(0) \approx 0.22\)–\(0.26\)) leads to substantial shell deformation and the development of secondary asymmetries. To highlight the influence of laser imprinting, the affected region is placed in the middle.  In contrast, when decreasing the imprinting level to $\frac{\delta h_{\text{proxy}}}{\sigma_{\text{tar}}(0)}  \approx 0.07$-$0.08$, the shell structure is virtually indistinguishable from the case with no laser imprinting. This confirms that laser imprinting has a negligible impact on performance degradation. Overall, for linear perturbations of medium‑to‑high order modes, the equivalent perturbation model can deepen our physical understanding of the evolution of $\delta h_{\text{proxy}}$ and $\sigma_\text{tar}(0)$ and serve as a simplified tool to optimize the implosion performance.    


\section{Conclusions}
In conclusion, this study develops an equivalent perturbation model for direct-drive ICF that quantifies laser imprinting as target surface perturbations. The model identifies a implosion sensitivity to laser imprinting $\frac{\delta h_\text{proxy}}{\delta h_\text{tar}(0)}= 0.1$. When $\frac{\delta h_\text{proxy}}{\delta h_\text{tar}(0)}\leq 0.1$, laser imprinting alters the nonlinear onset time and adiabat at the start of acceleration by less than 12\%, compared to simulations with $\delta h_\text{tar}(0)$ alone. Moreover, the implosion sensitivity to laser imprinting is supported by OMEGA experiments. These results indicate that joint control of multimode $\delta h_{\text{proxy}}$ and $\sigma_\text{tar}(0)$, improving target quality when $\frac{\delta h_\text{proxy}}{\sigma_\text{tar}(0)}\leq 0.1$ and laser smoothing otherwise, can support stable implosions for high-gain direct-drive inertial confinement fusion. Overall, for linear perturbations of medium‑to‑high modes, the  model can deepen our physical understanding of the evolution of $\delta h_{\text{proxy}}$ and $\sigma_\text{tar}(0)$ and serve as a simplified tool to optimize the implosion performance. The paper currently focus on the target surface perturbation and future work will extend the model to more internal target perturbations.


\begin{acknowledgments}
This work is supported by National Natural Science Foundation of China (Grant No. 12375242), and Science Challenge Project (Grant No. TZ2025014).
\end{acknowledgments}
\section*{Data Availability Statement}
Data available on request from the authors.
\section{Appendix}
\subsection{Simulations including one perturbation}
Table \ref{T1} lists simulation cases that satisfy $\frac{\delta h_\text{proxy}}{\delta h_\text{tar}(0)}\approx 1$ when $D_{ac}\approx 1$. These cases were simulated using the two-dimensional Eulerian radiation-hydrodynamic code FLASH \cite{fryxell2000flash} to validate Eq. (\ref{eq:deformation}).

A common setup was used for all simulations. A planar CH target with a density of 1 g/cm$^3$ is placed at Y = [0 $\mu$m, 40 $\mu$m]. The vertically incident laser employs third-harmonic and offers two types of pulse shapes. Pulse-A is a square pulse with a rise time of 0.1 ns and a peak intensity of 100 TW/cm$^2$.
Pulse-B is also a square pulse with a longer rise time of 0.5 ns and a peak intensity of 75 TW/cm$^2$. Compared to Pulse-A, Pulse-B is designed to enhance the imprinting of laser perturbations. The simulation domain has dimensions of ${L}_{\text{X}}=(1-4)\lambda$ in the $\text{X}$-direction and ${{L}_{\text{Y}}}= [-400\ \mu\text{m},\ 200\ \mu\text{m}]$ in the $\text{Y}$-direction, with a spatial resolution ranging from 0.13 $\mu$m to 0.52 $\mu$m. Here, $\lambda$ represents the single-mode perturbation wavelength applied to both the laser and the target. $\frac{\delta I_L}{I_L}(x)=\frac{\delta I_L}{I_L}\cos(2\pi x/\lambda)$ and $\delta h (x)= \delta h_\text{tar}(0) \cos(2\pi x/\lambda)$, where the amplitude $\delta h_\text{tar}(0)$ is either 2 $\mu$m or 4 $\mu$m.
 \begin{table}[h]
    \centering
    \begin{tabular}{cccccc||cccccc}
    \hline
pulse&  case& $\frac{\delta I_L}{I_L}$ 
&$t_\text{load}(\text{ns})$&$\lambda(\mu \text{m})$  &$\delta h_\text{proxy}(\mu \text{m})$& pulse&  case& $\frac{\delta I_L}{I_L}$ 
&$t_\text{load}(\text{ns})$&$\lambda(\mu \text{m})$  &$\delta h_\text{tar}(0)(\mu \text{m})$\\
        \cline{1-12}
         A&  1&  0.96&  0.2 &30&2& A&  5& /& / &30&2\\    
         A&  2& 0.7 & 0.3&50&2&A&  6& /& /&50&2\\
         A&  3& 0.64 & 0.5&100&4&A&  7& /& /&100&4\\        
         A&  4& 0.38 & 0.9&150&4&A&  8& /& /&150&4\\
         \cline{1-12}
         B&  9& 0.62 & 0.3&30&2&B&  13& /& /&30&2\\
         B&  10& 0.36 & 0.5&50&2&B&  14& /& /&50&2\\
         B&  11& 0.55 & 0.6&70&4&B&  15& /& /&70&4\\
         B&  12& 0.38 & 0.9&100&4&B&  16& /& /&100&4\\

\hline
    \end{tabular}
    \caption{Information of simulation cases with one kind of perturbation. Here, $t_\text{load}$ represents the imprinting time of laser perturbations, and it increases with $\lambda$. }
    \label{T1}
\end{table}

According to steady‑ablation theory, transitioning from a high‑entropy (pulse A) to a low‑entropy (pulse B) pulse design reduces the laser intensity \(I_L\). This decreases the conduction‑zone width (\(D_{ac} \sim I_L^{4/3}\)) and lowers the perturbed ablation velocity (\(\delta v_a \sim (\frac{\delta I_L}{I_L})^{1/3}I_L^{1/3}\)). The loading time \(t_\text{load}\), defined by the condition \(k D_{ac} = 1\), then scales as \(t_\text{load} \sim I_L^{-4/3}\lambda\). Consequently, $\delta h_\text{proxy}$ (\(\delta h_\text{proxy} \sim \delta v_a t_\text{load} \sim I_L^{-1}(\frac{\delta I_L}{I_L})^{1/3}\lambda\)) increases. This can been seen in Table \ref{T1}. Compared to pulse B, the same $\delta h_\text{proxy}$ and $\lambda$ correspond to a larger $\delta I_L/I_L$ under pulse A. For the same $\delta h_\text{proxy}$ and $I_L$, a smaller laser perturbation scale results in a shorter imprinting time onto the target, and thus a larger $\delta I_L/I_L$; conversely, a larger laser perturbation scale leads to a longer imprinting time, and thus a smaller $\delta I_L/I_L$. 
 \subsection{Simulations including two perturbations}
The cases in Table \ref{T2} feature both $\delta I_L/I_L$ and $\delta h_\text{tar}(0)$ ($\sigma_\text{tar}(0)$). These cases were simulated to investigate the influence of laser imprinting on the onset time of the nonlinear stage and on the adiabat, compared to cases with only $\delta h_\text{tar}(0)$ ($\sigma_\text{tar}(0)$). The common simulation setup is identical to that described for Table \ref{T1}. The multimode $\delta I_L/I_L$ is prescribed as:
$\frac{ \delta I_L}{I_L}(x)=\frac{ \delta I_L}{I_L}(\cos({2\pi x/30+\phi_1})+\cos({2\pi x/40+\phi_2})+\cos({2\pi x/60+\phi_3})+\cos({2\pi x/100+\phi_4}))$, where $x$ is in unit of $\mu$m. The multimode $\sigma_\text{tar}(0)$ is set as $\sigma_\text{tar}(0)=\text{rms}(\delta h(x))$, $ \delta h(x)=\left(100k^{-2}\sum\cos({2\pi /200*nx+\phi})(\mu \text{m})\right)$, where n is an integer ranging from 4 to 10, ${\phi }$ and $\phi_{1-4}$ are random phase uniformly distributed between zero and one.
\begin{table}[h]

    \centering
    \begin{tabular}{cccccc||cccccc}
    \hline
    \makecell{pulse} & \makecell{case} & \makecell{$\frac{\delta I_L}{I_L}$} 
    & \makecell{$t_\text{load}$\\($\text{ns}$)} & \makecell{$\lambda_\text{tar}(\mu\text{m})$\\$/\lambda_\text{las}(\mu\text{m})$} & \makecell{$\delta h_\text{tar}(0)$\\($\mu\text{m}$)} 
    & \makecell{pulse} & \makecell{case} & \makecell{$\frac{\delta I_L}{I_L}$} 
    & \makecell{$t_\text{load}$\\($\text{ns}$)} & \makecell{$\lambda_\text{tar}(\mu\text{m})$\\$/\lambda_\text{las}(\mu\text{m})$} & \makecell{$\delta h_\text{tar}(0)(\mu\text{m})$\\or $ \sigma_\text{tar}(0)(\mu\text{m})$} \\
    \cline{1-12}
    A & 17 & 0.0 & 0.3 & 50/50 & 4 & A & 30 & 0.3& 0.3& 50/50 & 2 \\    
    A & 18 & 0.1 & 0.3 & 50/50 & 4 & A & 31 & 0.4 & 0.3 & 50/50 & 2 \\
    A & 19 & 0.2 & 0.3 & 50/50 & 4 & A & 32 & 0.5& 0.3 & 50/50 & 2 \\        
    A & 20 & 0.3 & 0.3 & 50/50 & 4 & A & 33 & 0.6 &0.3 &50/50 & 2 \\
    A & 21 & 0.4 & 0.3 & 50/50 & 4 & A & 34 & 0.7 & 0.3 & 50/50 & 2 \\
    A & 22 & 0.5 & 0.3 & 50/50 & 4 & B & 35 & 0.0& 0.9 & multimode & 0.76 \\
    A & 23 & 0.6 & 0.3 & 50/50 & 4 & B & 36 & 0.05& 0.9 & multimode & 0.76 \\
    A & 24 & 0.7 & 0.3 & 50/50 & 4 & B & 37 & 0.07& 0.9 & multimode & 0.76 \\
    A & 25 & 0.3& 0.2& 50/30 & 4& B & 38 & 0.1& 0.9& multimode & 0.76\\
     A & 26 & 0.3 & 0.5 & 50/100 & 4& B & 39 & 0.15& 0.9& multimode & 0.76\\
     A & 27 & 0.3& 0.9 & 50/200 & 4& B & 40 & 0.2& 0.9& multimode & 0.76\\
     A & 28 & 0.1 &0.3 &50/50 & 2& B & 41 & 0.3& 0.9& multimode & 0.76\\
     A & 29 & 0.2& 0.3& 50/50 & 2 & B & 42 & 0.4& 0.9& multimode & 0.76\\
    \hline
    \end{tabular}
    \caption{The information of simulation cases with both perturbations.}
    \label{T2}
\end{table}

\begin{figure}[h]
  \includegraphics[width=0.5 \linewidth]{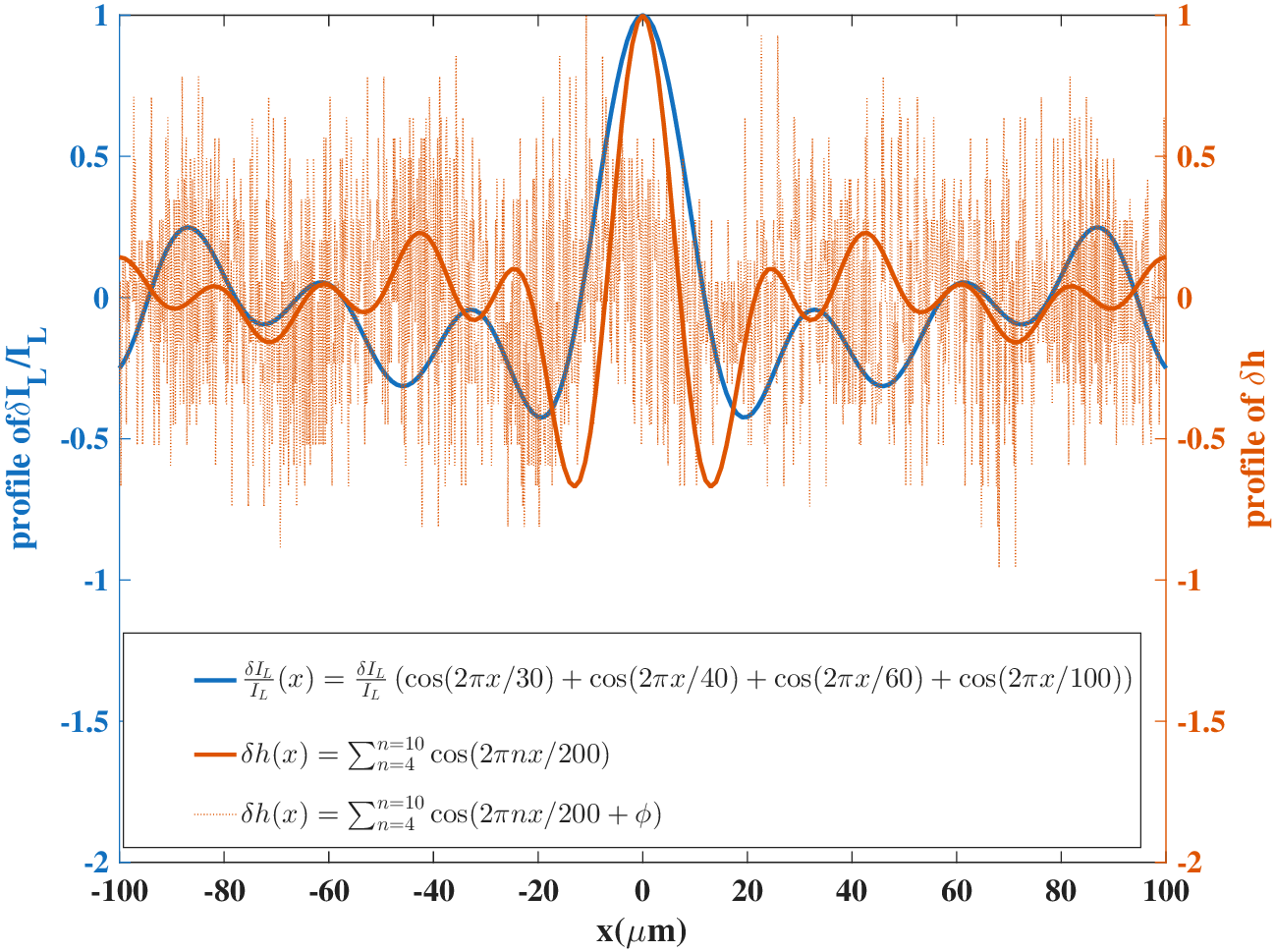}%
 \caption{\label{FIG. 7} Profiles of multimode $\delta h$ and $\frac{\delta I_L}{I_L}$. The orange dotted line is from simulation.} 
 \end{figure}

The profiles of multimode $\delta h$ and $\frac{\delta I_L}{I_L}$ are shown in Fig. \ref{FIG. 7}. In the absence of random seeds, both \(\delta h\) and \(\frac{\delta I_L}{I_L}\) have their maximum perturbations at \(x = 0\,\mu{m}\), where \(\frac{\delta I_L}{I_L}\) is most influential on the shell. When \(\frac{\delta h_\text{proxy}}{\sigma_\text{tar}(0)} = 0.22\sim 0.26\), a dip due to \(\frac{\delta I_L}{I_L}\) appears at \(x = 0\,\mu{m}\).

\bibliography{main}

\end{document}